\documentclass[fleqn,usenatbib]{mnras}

\usepackage{newtxtext,newtxmath}

\usepackage[T1]{fontenc}

\DeclareRobustCommand{\VAN}[3]{#2}
\let\VANthebibliography\thebibliography
\def\thebibliography{\DeclareRobustCommand{\VAN}[3]{##3}\VANthebibliography}

\usepackage{graphicx}	
\usepackage{amsmath}
\usepackage{amssymb}	
\usepackage{longtable}
\usepackage{threeparttable}
\usepackage{dcolumn}
\usepackage{multirow}
\usepackage{booktabs}
\usepackage{array}

\title[HS 2325+8205]{First discovery of QPOs in the dwarf nova HS 2325+8205 based on TESS photometry}

\author[Sun et al.]{
Qi-Bin Sun$^{1}$, 
Sheng-Bang Qian$^{1,2}$\thanks{E-mail: qsb@ynao.ac.cn},
Li-Ying Zhu$^{1}$,
Ai-Jun Dong$^{2}$,
Qi-Jun Zhi$^{2}$,
Wen-Ping Liao$^{1}$,
\newauthor{
Er-Gang Zhao$^{1}$,
Zhong-Tao Han$^{1}$,
Wei Liu$^{1}$,
Lei Zang$^{1}$,
Fu-Xing Li$^{1}$,
and Xiang-Dong Shi$^{1}$}
\\
$^{1}$Yunnan Observatories, Chinese Academy of Sciences (CAS), P.O. Box 110, Kunming, 650011, China\\
$^{2}$School of Physics and Electronic Science, Guizhou Normal University, Guiyang, 550001, China}

\date{Accepted 2022 November 07. Received 2022 October 06; in original form 2022 August 30 
}

\pubyear{2015}

\begin{document}
\label{firstpage}
\pagerange{\pageref{firstpage}--\pageref{lastpage}}
\maketitle


\begin{abstract}
HS 2325+8205 is a long-period eclipsing dwarf nova with an orbital period above the period gap ($P_{orb} > 3 h$) and is reported to be a Z Cam-type dwarf nova.
Based on the photometry of the Transiting Exoplanet Survey Satellite (TESS), the light variation and the quasi-periodic oscillation (QPOs) of HS 2325+8205 are studied.
Using Continuous Wavelet Transform (CWT), Lomb–Scargle Periodogram (LSP), and sine fitting methods, we find for the first time that there is a QPOs of $\sim 2160$s in the long outburst top light curves of HS 2325+8205. Moreover, we find that the oscillation intensity of the QPOs of HS 2325+8205 is related to the orbital phase, and the intensity in orbital phases 0.5-0.9 are stronger than in orbital phases 0.1-0.5.
Therefore, the relationship between the oscillation intensity of QPOs and the orbital phase may become a research window for the origin of QPOs.
In addition, we use the LSP to correct the orbital period of HS 2325+8205 as 0.19433475(6) d. 

\end{abstract}




\section{Introduction} \label{sec:intro}

Dwarf novae (DNe) are a subtype of cataclysmic variable stars (CVs), a semi-detached close binary system consisting of a white dwarf (WD; primary) and a late-type star (secondary). Secondary matter fills the Roche lobe, and WD accreted matter from the secondary star through the inner Lagrangian point \citep{warner1995cat}. 
The transferred material from the secondary star forms a bright collision zone (hot spots) with the edge of the accretion disk \citep{warner1995cat}.
Compared with other CVs, DNe have intermittent outbursts with small amplitudes, and the luminosity increases by about 2-8 magnitudes from a few days to a few weeks. 
According to the different characteristics of the outburst, DNe can be divided into three main subtypes: Z Cam, SU UMa, and U Gem. The outburst can be explained by the thermal limit-cycle instability in the accretion disk (DIM, \citealp{lasota2001disc}). 
The orbital evolution of CVs are driven by angular momentum loss (AML). The remarkable feature is the obvious lack of CVs with a period of about 2-3 hours, which is the famous orbital period gap \citep{knigge2006donor}. The standard model explains the period gap by assuming a magnetic interruption mechanism, the magnetic braking (MB, \citealp{rappaport1983new}) controls the AML above the period gap ($P_{orb} > 3 h$), while the gravitational radiation (GR, \citealp{paczynski1981gravitational}) drives the evolution of CVs in a shorter period ($P_{orb} < 2 h$).
The observed-minus-calculated (O–C) method of mid-eclipse times is an effective method to study the orbital period variation and is widely used to study the orbital period evolution of eclipsed binary stars. A large number of studies have shown that orbital cyclic variation exists in a large number of CVs (e.g., V2051 Oph, \citealp{qian2015long}; EX Dra, \citealp{han2017double}; U Gem, \citealp{dai2009orbital}; EM Cyg, \citealp{liu2021quasi}). 
It is generally accepted that DNe have three types of rapid oscillations \citep{warner2004rapid}, DNe oscillations (DNOs), long-period DNe oscillations (lp-DNOs), and quasi-periodic oscillations (QPOs). QPOs was first proposed by \citet{patterson1977rapid}, and the period is longer than DNOs and lp-DNOs, ranging from hundreds of seconds to thousands of seconds.

HS 2325+8205 was classified as a CV candidate by \citet{morgenroth193677}. It was classified as CV by \citet{aungwerojwit2005hs} in the Hamburg Quasar Survey (HQS, \citealp{hagen1995hamburg}). \citet{shears2011hs2325+} observed 44 outbursts of HS 2325+8205 from 2007 to 2009, of which 33 occurred in a short period, the duration of the outburst was less than 7 days, the brightness was less than 14 mag, and the duration of the long outburst was greater than 9 days, the brightness was greater than 13.9 mag.
\citet{pyrzas2012hs} made photometry and spectroscopic observations on HS 2325+8205 from 2003 to 2007 and obtained many mid-eclipse times. 
Because of the short recurrence time $(12 - 14^d)$, they classified HS 2325+8205 as a Z Cam-type DN.
\citet{hardy2017hunting} also carried out photometry analysis on HS 2325+8205 and obtained three mid-eclipse times. The QPOs of HS 2325+8205 have not been studied due to the lack of observation. 
Based on the TESS photometry, we conducted an in-depth study on the light variations and QPOs.

In this paper, we investigated the data of HS 2325+8205 from TESS to study the light variations and the QPOs. 
This paper is structured as follows. Section 2 briefly describes the data of HS 2325+8205 observed by TESS.
Section 3 presents the characteristics of light variation and the detection of QPOs. In section 4, We discuss the physics of the origin of QPOs. Section 5 is the summary.


\section{TESS Photometry} \label{sec:style}

We used the data released by the space telescope Transiting Exoplanet Survey Satellite (TESS, \citealp{Ricker2015journal}), which was launched in 2018, and these were generated by the TESS Science Processing Operations Center (SPOC, \citealp{Jenkins2016}) of NASA Ames Research Center, which is managed by the TESS Payload Operations Center (POC) of Massachusetts Institute of Technology (MIT). The wavelength of TESS's observation range from 600 to 1000 nm, covering 2100 skies per pixel \citep{Ricker2015journal}. In TESS's observation, the sky is divided into several sectors, and each sector observation is about one month. The sector observation has a long cadence (30 minutes) mode and a short cadence (2 minutes) mode. 

\begin{figure*}
\includegraphics[width=1.5\columnwidth]{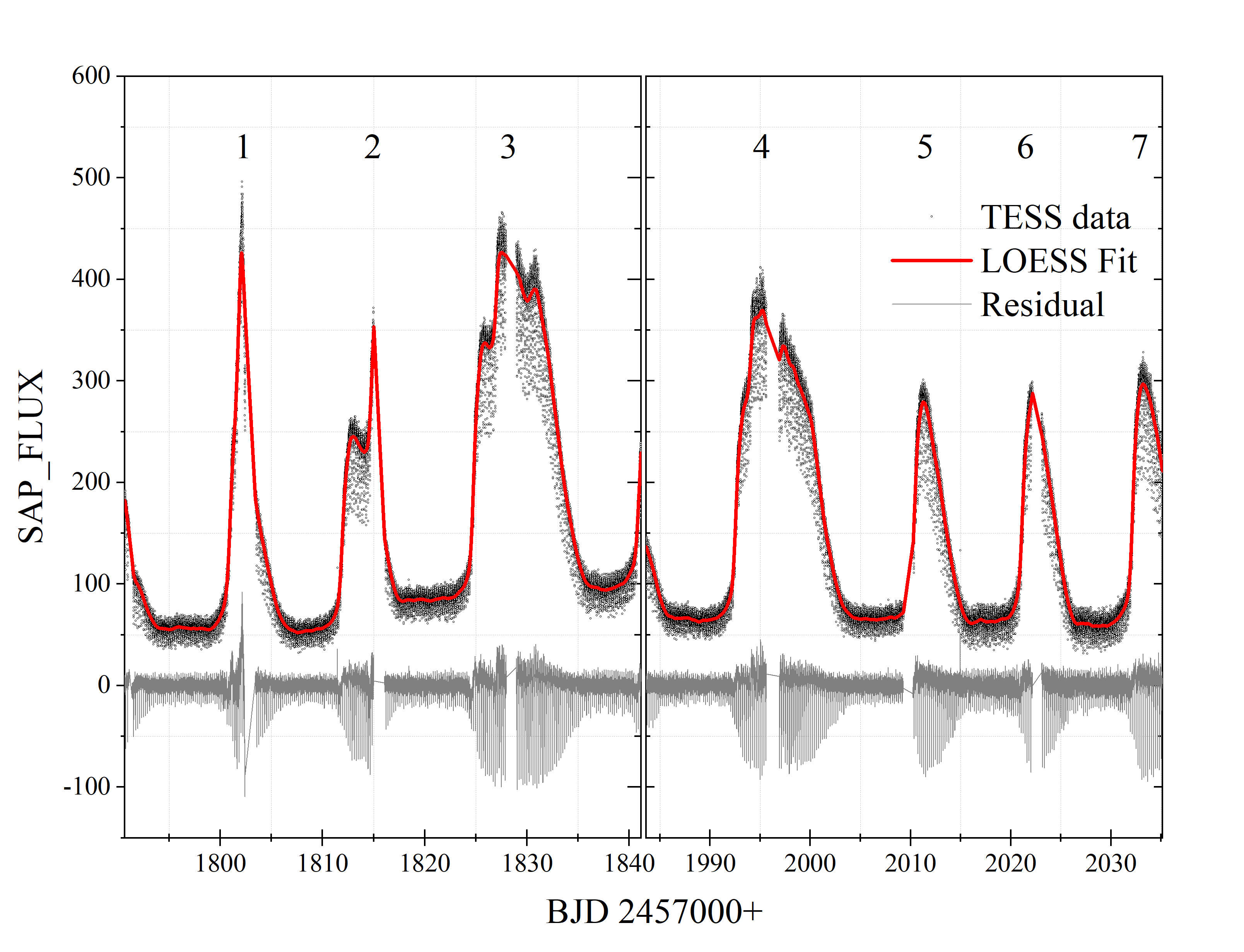}
\caption{Light curves of HS 2325+8205. The solid red line is the LOESS fit, and the solid gray line is the residual. \label{fig:light curve}}
\end{figure*}

HS 2325+8205 was observed by TESS in sectors 18, 19, 25, and 26, respectively. 
The observation time of sector 18 is from November 3, 2019 (MJD 58790) to November 27, 2019 (MJD 58814), sector 19 is from November 28, 2019 (MJD 58815) to December 23, 2019 (MJD 58840), and sector 25 is from May 14, 2020 (MJD 58983) to June 8, 2020 (MJD 59008), Sector 26 was observed from June 9, 2020 (MJD 59009) to July 4, 2020 (MJD 59034).
Each sector has a data gap of about one day due to data transmission. The total time of the four observations is about 94 days, with 68,685 data points, and the exposure time is 120s in a short cadence mode (2 minutes). A total of seven outbursts were observed in the four observations, which provided a lot of data support for the study of HS 2325+8205 (see Figure\ref{fig:light curve}).

\section{Data Analysis and Results} \label{sec:Analysis}

\subsection{Eclipse and Period} \label{subsec:O-C}

\begin{figure*}
\includegraphics[width=1.5\columnwidth]{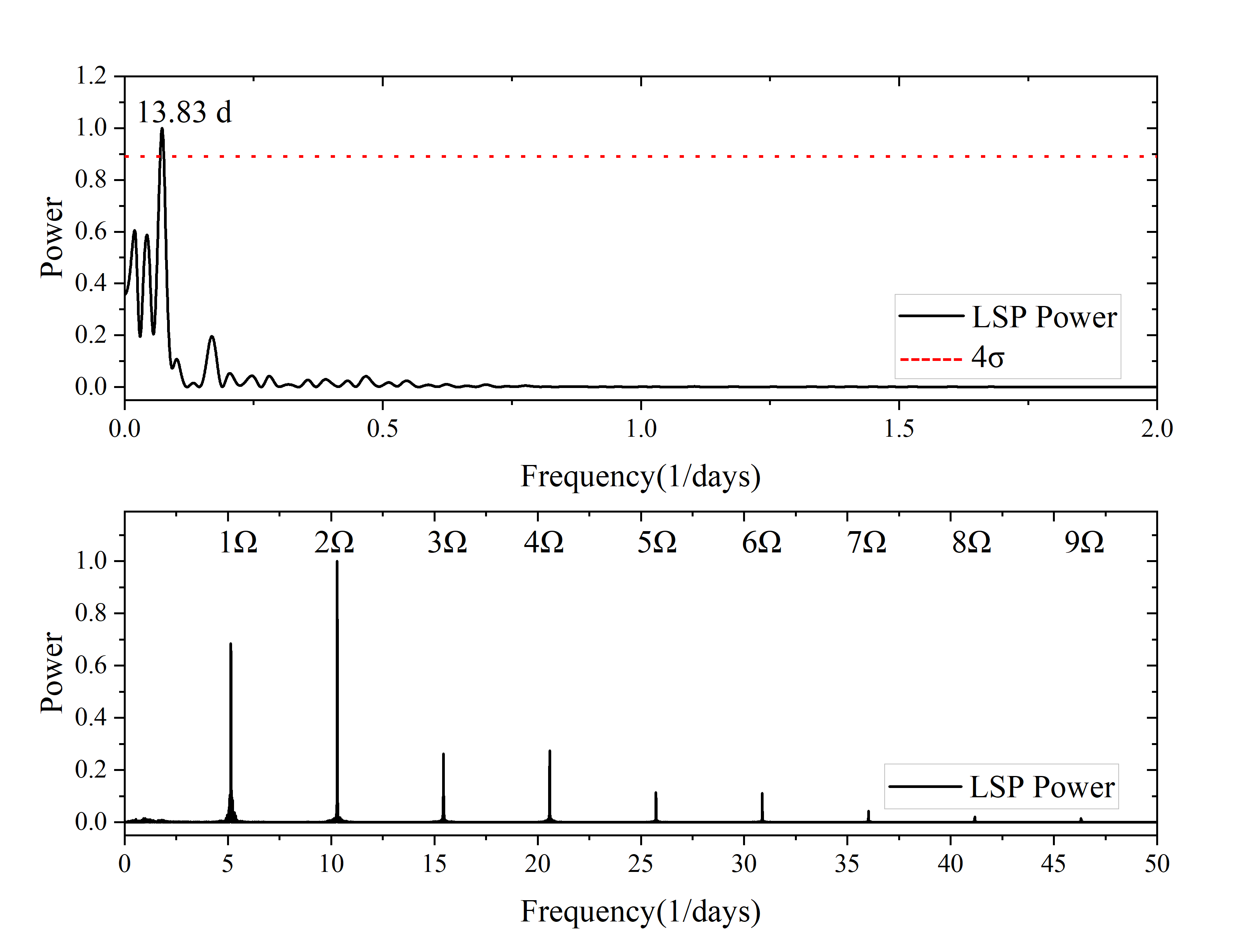}
\caption{LSP of HS 2325+8205. The top panel is the LSP of the recurrence period, and the red dotted line in the figure is 4 times the standard deviation (4$\sigma$). The bottom panel is the LSP of the orbital period, and we use the symbol $\Omega$ for the frequency of the orbital period and its harmonics. \label{fig:LSP}}
\end{figure*}

TESS data of HS 2325+8205 contains many eclipse curves, which provide excellent data support for us to study the orbital period. 
Firstly, we used locally weighted regression (LOESS; \citealp{cleveland1979robust}) with a smoothing span of 0.1 to smooth the light curves.
Then, the smooth curves were subtracted from the original light curves to obtain the light curves with the trend of outburst removed (see Figure\ref{fig:light curve}). 
Finally, the Lomb–Scargle Periodogram (LSP; \citealp{lomb1976least}) was used to detect periodic signals. Through LSP we find that the orbital period of HS 2325+8205 is 0.19433475(6)d ($\Omega=5.145760\pm1.56\times10^{-6}$$days^{-1}$, see Figure \ref{fig:LSP}). The uncertainty is based on the half-width at half maximum (HWHM) Gaussian fitting centered on the significant power. 

We used the Gaussian function to fit the eclipse provided by TESS, and the code of Gaussian fit comes from \cite{Fang2020ApJ}. Due to the low resolution of TESS data, we used the orbital phase width of 0.1 centering on the eclipse for fitting.
491 new mid-eclipse times were obtained (see Supplementary Material). 
Taking the first mid-eclipse times of quiescence as the initial epoch and taking 0.19433475(6)d as the orbital period, a new ephemeris is obtained:

\begin{equation}\label{eq:ephemeris}
T_{mid}=BJD 2458793.6762(6) + 0.19433475(6) \times E  
\end{equation}

\begin{figure}
\includegraphics[width=\columnwidth]{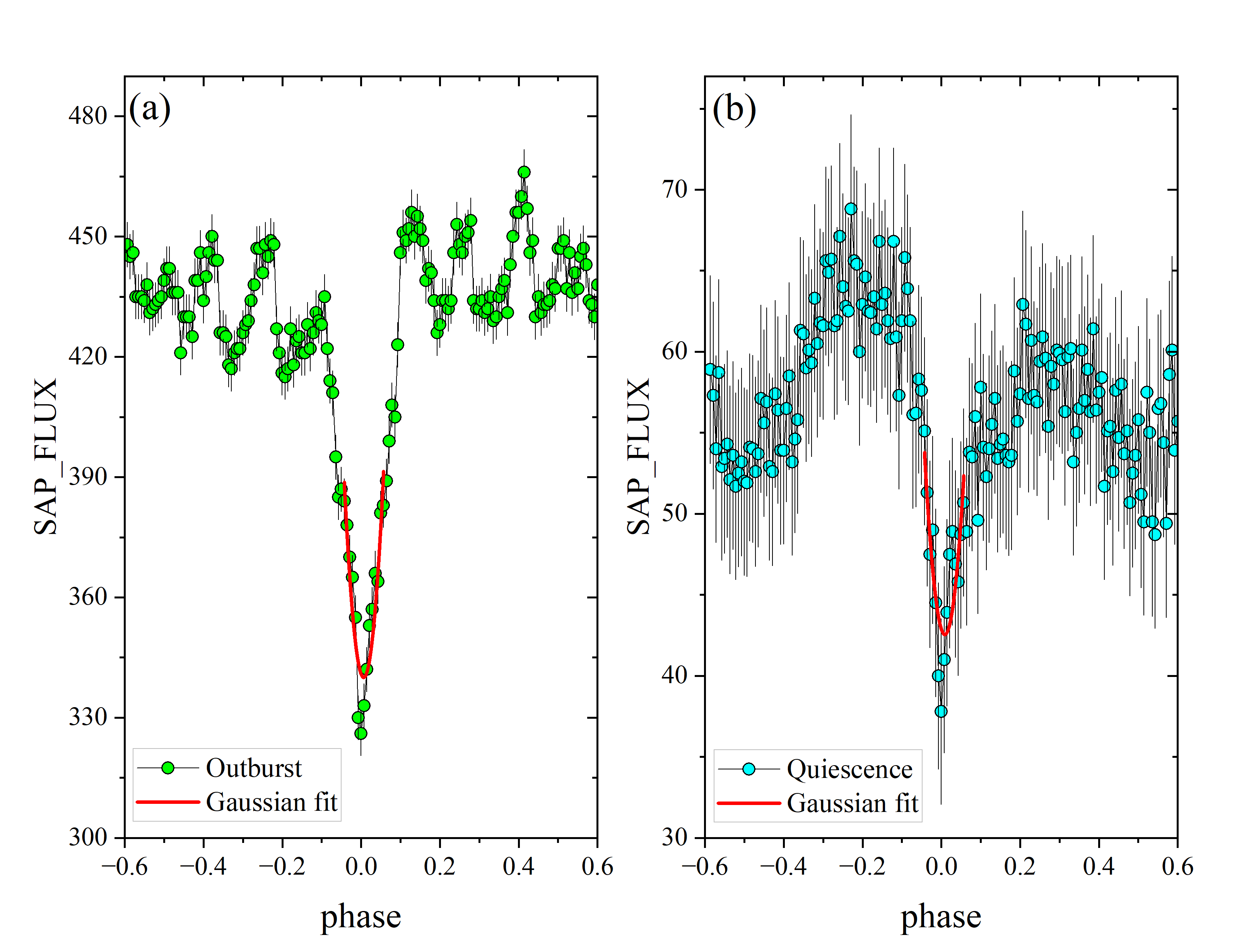}
\caption{The example of using Gaussian fit to obtain the minima. (a) and (b) are in outburst (E = 174) and quiescence (E = 15) respectively. The solid red line is the Gaussian fit of the eclipse profile with phase widths of 0.1.\label{fig:minima}}
\end{figure}

\label{key}\subsection{Outbursts and double humps} \label{subsec:outbursts}

\begin{figure}
\includegraphics[width=\columnwidth]{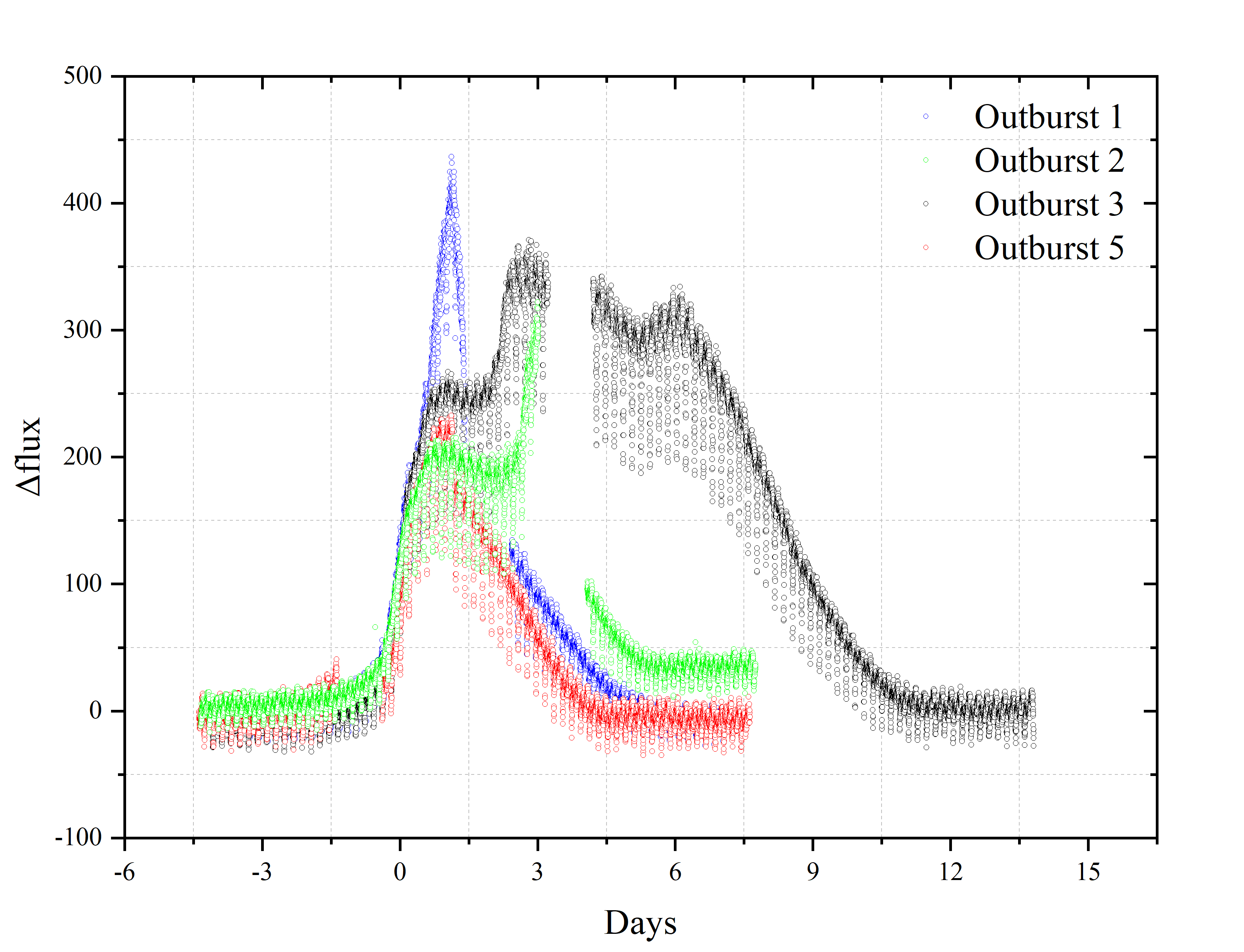}
\caption{Comparison of four different outburst curves. \label{fig:three outbursts}}
\end{figure}

TESS observed seven outbursts of HS 2325+8205 from four different sectors (18, 19, 25, and 26).
All outbursts are numbered 1-7 by us in the order of outbursts (see Figure\ref{fig:light curve}).
Depending on the duration of the outbursts, outbursts 1, 2, 5, and 6 are short outbursts with a duration of $\sim 7^d$, while outbursts 3 and 4 are long outbursts with a duration of $\sim 13^d$. 
Long outbursts are longer than $\sim 9^d$ observed by \citet{shears2011hs2325+}.
Four different outburst types can be distinguished by the amplitude and duration of the outbursts (see Figure \ref{fig:three outbursts}). Outburst 1 shows a short duration ($\sim 7^d$) but a large amplitude. Outburst 3 has a long duration ($\sim 13^d$) and a large amplitude, with the mid-brightness interval. Outburst 5 has a short duration ($\sim 7^d$) and a small amplitude. Outburst 2 has a short outburst duration with a mid-brightness interval. In other TESS observations, it can be found that the data will be abnormal near the end of an observation period, so the authenticity of outburst 1 and outburst 2 needs to be verified. 

The LSP was used to detect the recurrence time of the outburst and obtained that the recurrence time was $\sim 13.83^d$(see Figure \ref{fig:LSP}). 
Based on the Ritter and Kolb catalog \citep{ritter2003catalogue}, \citet{pyrzas2012hs} made statistics on 22 Z Cam-type DNe and found that their occurrence time is mainly distributed in $11-18^d$, which is consistent with the recurrence time of HS 2325+8205 observed by TESS. It is further confirmed that HS 2325+8205 is a Z Cam-type DN in the recurrence time, but the standstill of the typical characteristic of Z Cam-type DNe still needs to be verified by the observation.

We folded all the quiescent data of HS 2325+8205 using 0.19433475(6) d and found the peculiar double humps similar to XY Ari at phases 0.2-0.4 and 0.7-0.9, which can be explained by ellipsoidal modulation (\citealp{warner2003cataclysmic}).
According to the ellipsoidal modulation, the distorted Roche-lobe shape of the secondary star faces us with the maximum area at orbital phases 0.25 and 0.75, showing double humps on the light curve. During the orbital phases 0.0-0.5, the secondary star moves away from us, and during the orbital phases 0.5-1.0, it moves closer to us. Therefore, the brightness of the secondary star in the orbital phases 0.5-1.0 is brighter than that in the orbital phases 0.0-0.5, and the darkest occurs in the orbital phases $\sim 0.5$ of the secondary star eclipse.

\begin{figure}
\includegraphics[width=\columnwidth]{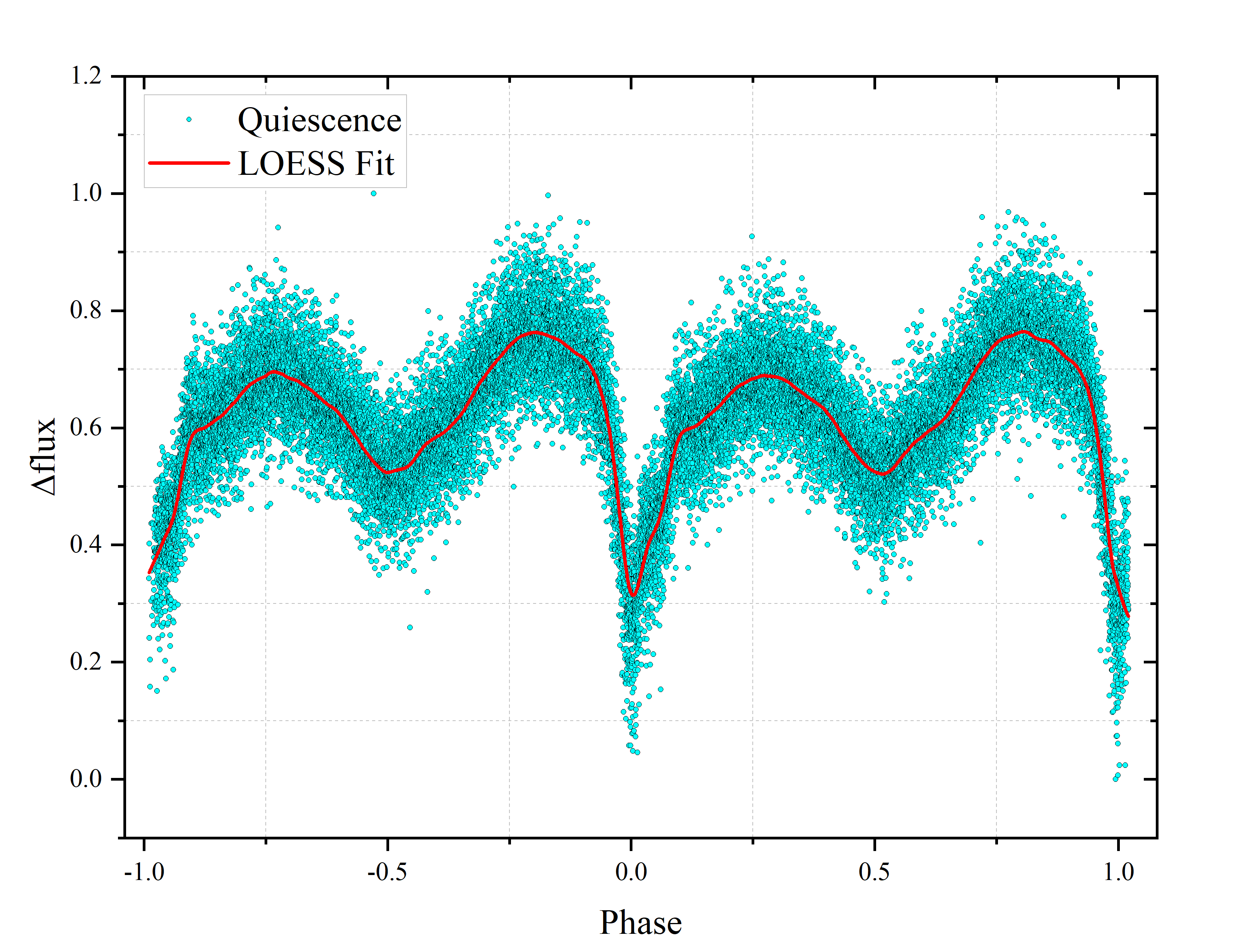}
\caption{Phase-folded light curve of HS 2325+8205 during quiescence. The solid red line is the LOESS fit, and the folding period is 0.19433475(6) d. \label{fig:QPhase}}
\end{figure}

\subsection{QPOs} \label{subsec:QPOs}

The current research shows that QPOs are widely present in CVs with high mass transfer rates, such as nova-like variables and DNe in outburst, and the QPOs appearing in DNe are most significant in the peak of outburst \citep{warner2004rapid,robinson1979quasi}.
By analyzing the TESS data of HS 2325+8205, we found that the out-of-eclipse curves had apparent oscillations at the peak of the long outburst of HS 2325+8205. 

\begin{figure}
\includegraphics[width=\columnwidth]{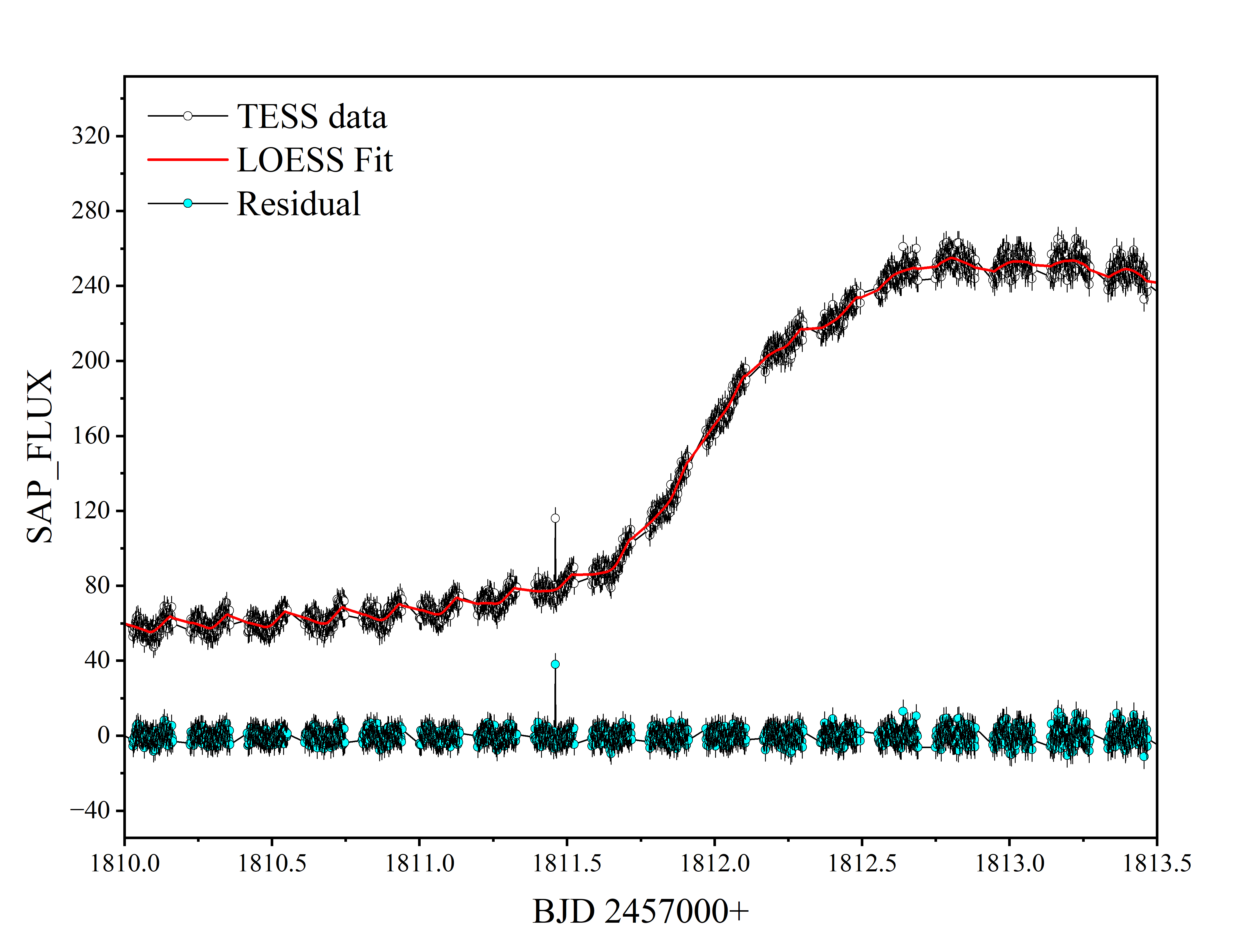}
\caption{Example of removing the trend of outburst and eclipse from the light curve. The data gap is the removed eclipse. \label{fig:example}}
\end{figure}

We first remove the eclipse part of the light curves with a new epoch \ref{eq:ephemeris}, then use LOESS with a smoothing span of 0.003 to fit all curves, to exclude the interference of double humps, and finally, the smooth curves were subtracted from the original light curves to obtain the out-of-eclipse curves with the outburst and the eclipse trend removed (see Figure \ref{fig:example}).

\begin{figure}
\includegraphics[width=\columnwidth]{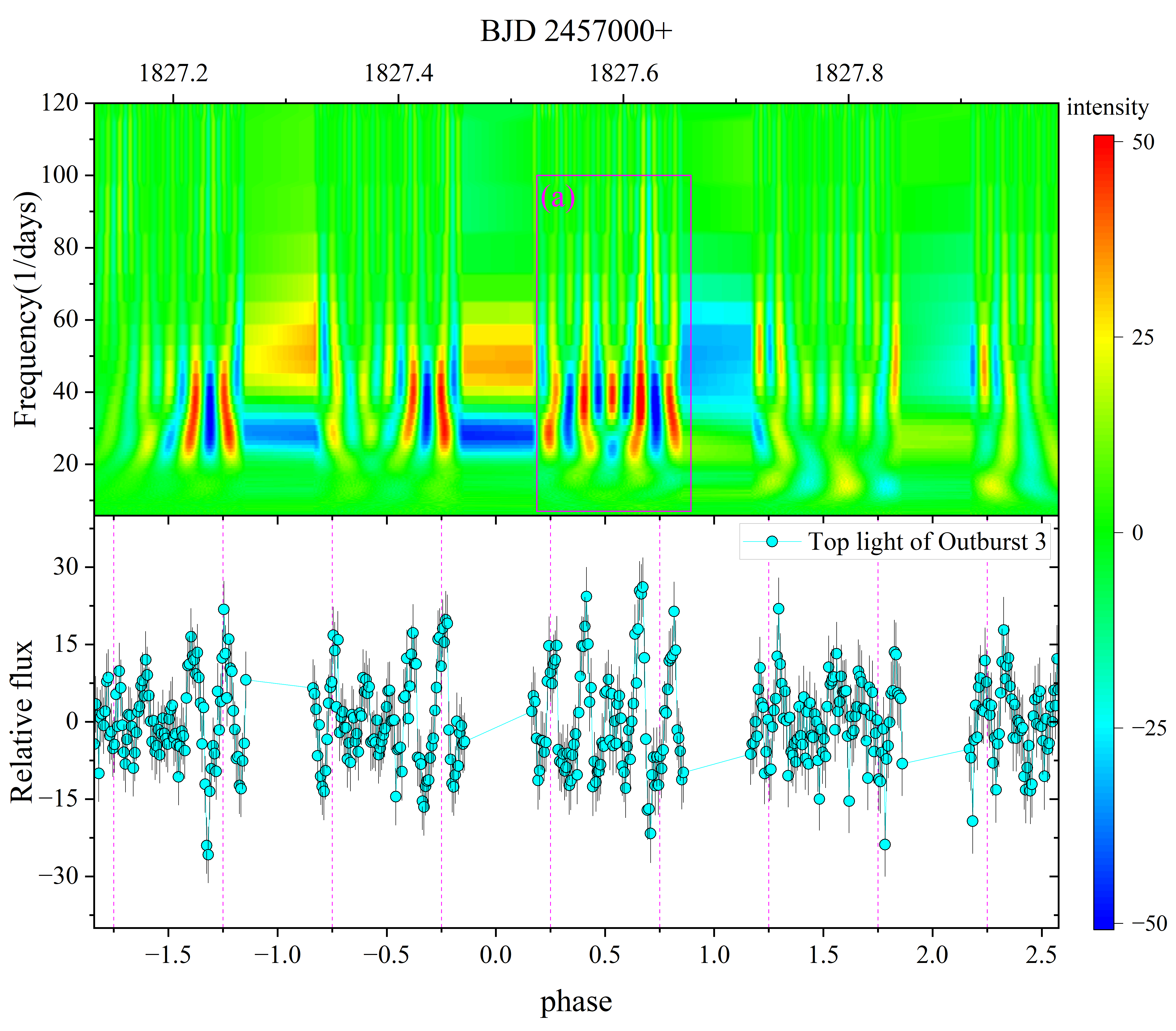}
\caption{CWT 2D power spectrum of the top curve of outburst 3 and its corresponding light curves. (a): The rectangular box is where the intensity of QPOs is greatest, and the abscissa is processed to the orbital phase. \label{fig:cwtlight3}}
\end{figure}

\begin{figure}
\includegraphics[width=\columnwidth]{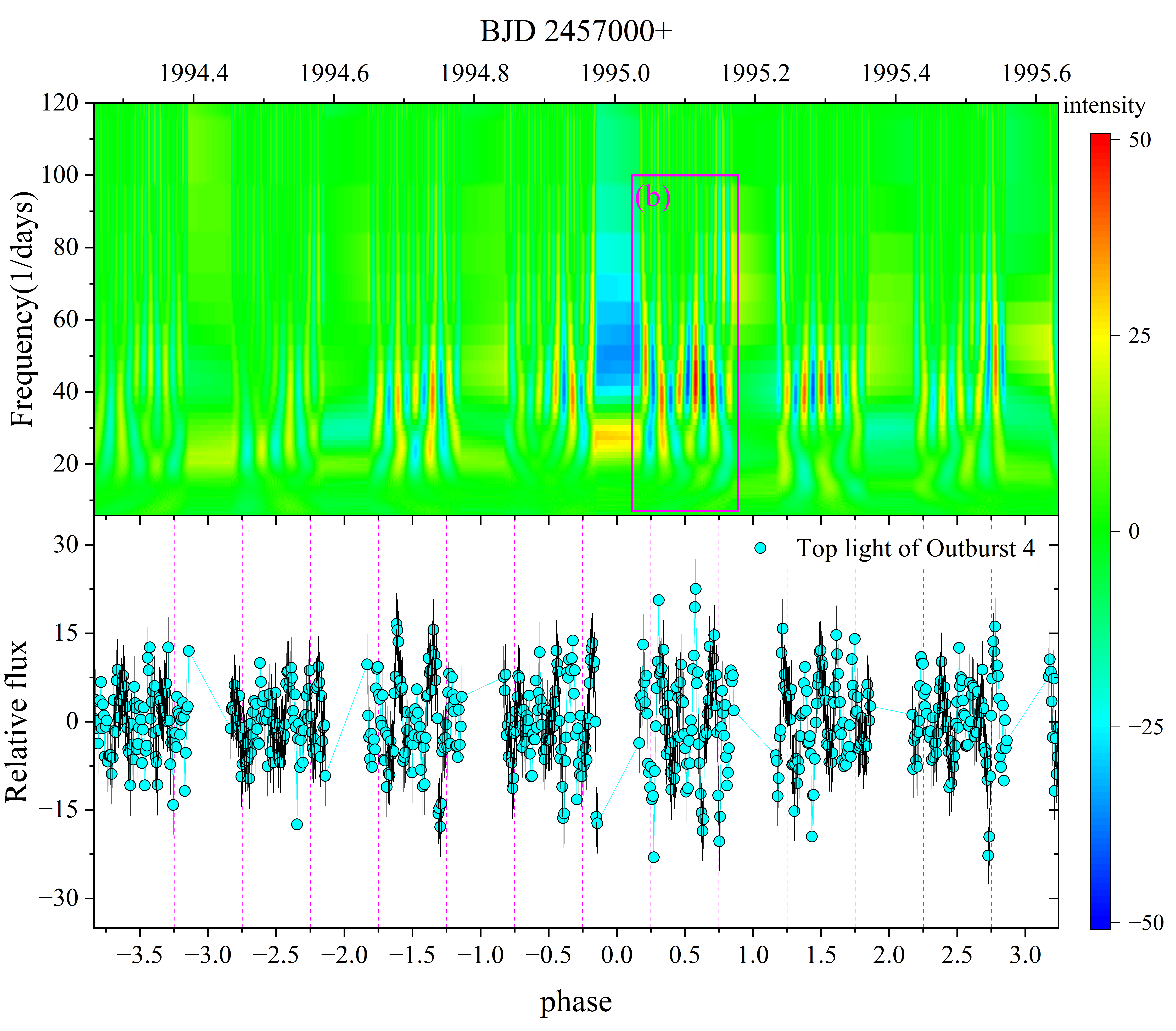}
\caption{CWT 2D power spectrum of the top curves of outburst 4 and its corresponding light curves. (b): The selected part is where the intensity of QPOs is greatest\label{fig:cwtlight4}}
\end{figure}

\begin{figure}
\includegraphics[width=\columnwidth]{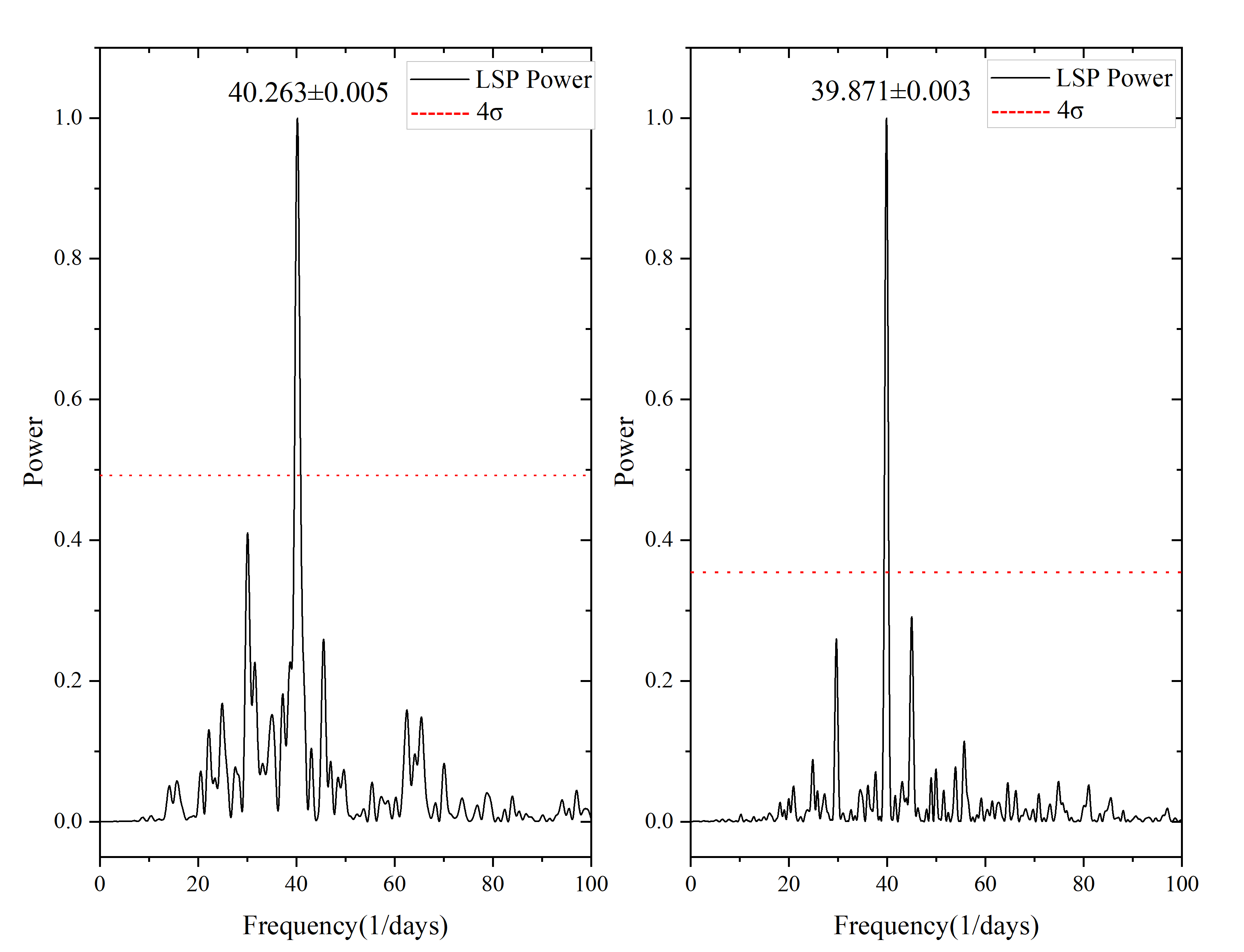}
\caption{The LSP of the top curves of outburst 3 and outburst 4. The left belongs to outburst 3, and the right belongs to outburst 4. The red dotted line is 4 times the standard deviation (4$\sigma$) of LSP.\label{fig:lsp34}}
\end{figure}

We analyzed all out-of-eclipse curves using a Continuous Wavelet Transform (CWT; \citealp{polikar1996wavelet}). The two-dimensional (2D) power spectrum results of the CWT show that there is a QPOs with frequency of $\sim 40$ $days^{-1}$ at the top of long outburst (see Figures \ref{fig:cwtlight3} and \ref{fig:cwtlight4}). Furthermore, we obtained the same results for the long outburst top curves using LSP (see Figure \ref{fig:lsp34}), which further confirms that the QPOs of HS 2325+8205 is $\sim 2160s$.

\begin{figure}
\includegraphics[width=\columnwidth]{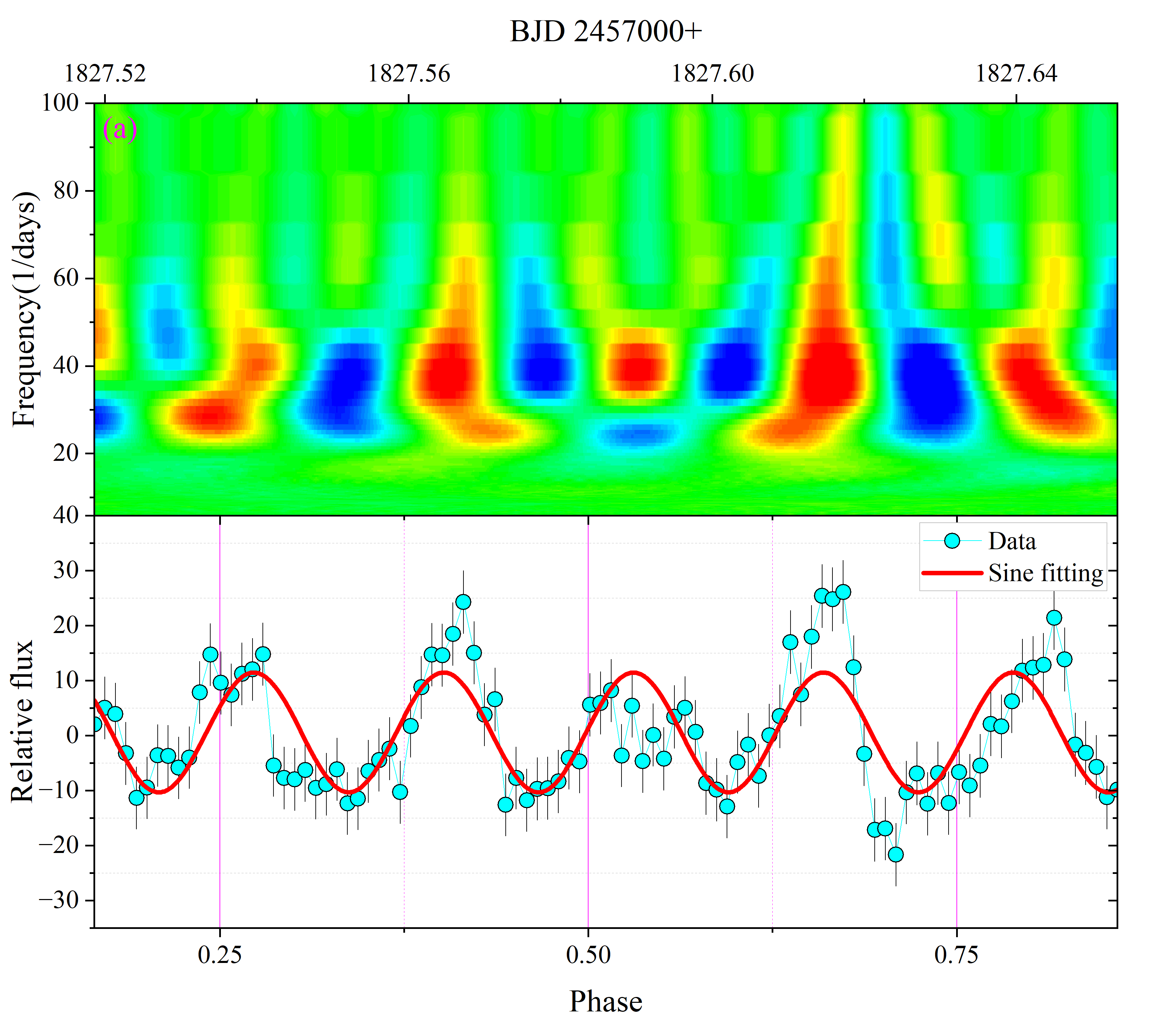}
\caption{The enlarged view of the area selected by the red rectangle (a) in Figure \ref{fig:cwtlight3} and its corresponding light curves. The solid red line in the bottom panel is the sine fit (the best fitting is $40.13\pm0.46$ $days^{-1}$), and the abscissa is processed to the orbital phase.\label{fig:CWTtop1}}
\end{figure}

\begin{figure}
\includegraphics[width=\columnwidth]{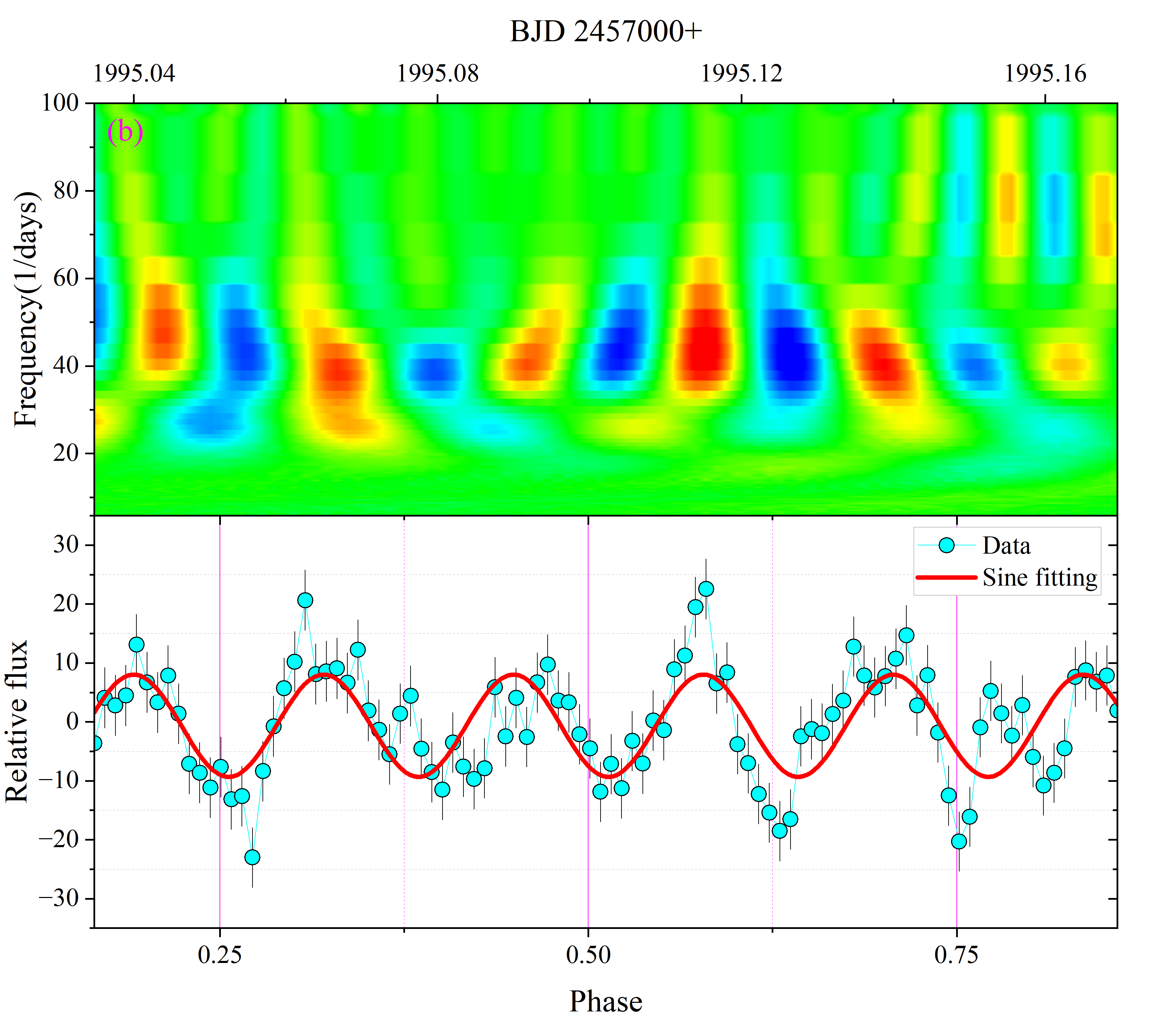}
\caption{The enlarged view of the area selected by the red rectangle (b) in Figure \ref{fig:cwtlight4} and its corresponding light curves. the best sine fitting is $ 39.79\pm0.39$ $days^{-1}$.\label{fig:CWTtop2}}
\end{figure}

Through the 2D power spectrum, we found that the oscillation intensity of QPOs is related to the orbital phase. We select the places with the strongest oscillation in outburst 3 and outburst 4, respectively, for analysis. We calculated the orbital phase corresponding to the light curves and performed sinusoidal fitting. By comparing the 2D power spectrum and the light curves, we found that the intensity of QPOs in orbital phases 0.5-0.9 are stronger than that in orbital phases 0.1-0.5, which is the weakest at $\sim 0.5$ (see Figures \ref{fig:CWTtop1} and \ref{fig:CWTtop2}). 
The intensity of QPOs in orbital phases 0.5-0.9 are stronger than in orbital phases 0.1-0.5 and also in other out-of-eclipse curves with significant oscillations (see Figures \ref{fig:cwtlight3} and \ref{fig:cwtlight4}), which may be ideal probes to study the origin of QPOs.

\begin{figure*}
\includegraphics[width=1.4\columnwidth]{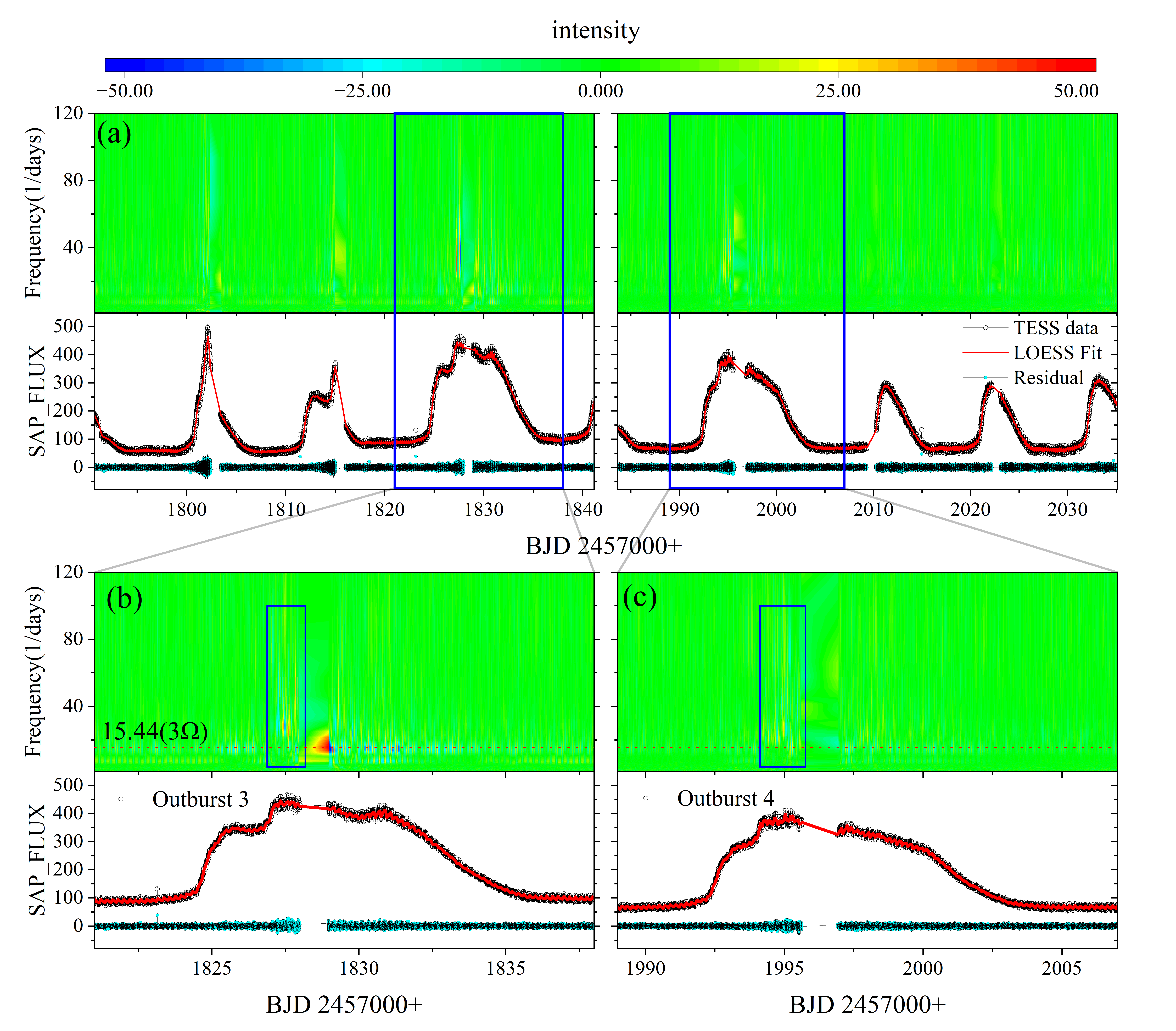}
\caption{CWT 2D power spectrum and their corresponding light curves. (a): All CWT 2D power spectrum and their corresponding light curves; (b) and (c) are the CWT 2D power spectrum and their corresponding light curves of Outburst 3 and Outburst 4. The rectangular box is the top light curve. The red dotted lines are the harmonics of the orbital period (3$\Omega$$\approx15.44 days^{-1}$). \label{fig:out34}}
\end{figure*}

\section{DISCUSSION} \label{sec:DISCUSSION}

\citet{warner2004rapid} suggested that there are at least three types of QPOs after statistical analysis of about 50 CVs. The first one is DNO-related QPOs, which coexist with DNOs and oscillate for several hundred seconds. The second is the IP-related QPOs, whose oscillation period is about 1000 s. Its generation is related to the structure of intermediate polars (IPs) \citep{patterson2002superhumps}. Except for DNO-related QPOs and IP-related QPOs, the remaining QPOs are the third (Other QPOs;
\citealp{warner2004rapid}, \citealp{scepi2021qpos}). 
The current study proposes several different explanations for the origin of QPOs, such as the internal oscillation of the accretion disk, the reprocessing of light by orbital blobs \citep{Patter1979Rapid}, the radial oscillation of the accretion disk \citep{okuda1992disc}, the interruption and reprocessing of traveling waves in the inner edge of accretion disk \citep{lubow1993wave}, and the influence of the magnetic field of the WD on the accretion disk \citep{paczynski1978phenomenological}.
QPOs widely exist in CVs with high mass transfer rates, such as nova-like variables and DNe in outbursts, and another consensus is that QPOs are closely related to accretion disks.

Through the analysis of the section \ref{subsec:QPOs}, we found that there are QPOs with oscillation periods of $ \sim 2160$ s in HS 2325+8205. Based on the classification of \citet{warner2004rapid}, it can be known that the rapid oscillation of $ \sim 2160$ belongs to Other QPOs, and there is no relevant theoretical model to describe it. 
In addition, we found that the intensity of QPOs is related to the orbital phase. 
The intensity of QPOs in orbital phases 0.5-0.9 are stronger than in orbital phases 0.1-0.5.
In the most significant light curve of QPOs, the intensity is the weakest at the orbital phase $\sim 0.5$.
Current research suggests that the light sources of the orbital signal mainly consist of three components:

(1) the irradiation of the secondary star by the accretion disk and/or the WD \citep{kimura2020thermal};

(2) the ellipsoidal modulation caused by the motion of the secondary star reaches its peak at the orbital phases 0.25 and 0.75 \citep{warner2003cataclysmic,kimura2020thermal};

(3) the orbital hump caused by hot spots mainly exists in the orbital phase of 0.7-0.9 during quiescence, and the outburst is not significant \citep{warner1995cat,Dous1994observations}.

(1) the irradiation of the secondary depends on the disk geometry and the disk luminosity; if the accretion disk is tilted, the surface of the secondary accretion disk will receive more radiation than in other cases because the internal hot part of the accretion disk is easily exposed to the surface of the secondary. At orbital phase 0.5, the irradiated area of the secondary star we observed is the largest, and the brightness reaches its peak. Negative superhumps are the most conspicuous sign of the tilted disk in CVs (e.g., \citealp{bonnet1985continuum}, \citealp{patterson19931991}, \citealp{patterson1999permanent}, \citealp{harvey1995superhumps}). 
During the analysis of the orbital period of HS 2325+8205, we did not find periodic signals shorter than the orbital period by about a few percent, so we will not consider (1).

Although the intensity variation of QPOs is similar to the phase of the double humps during quiescence, the QPOs exist only during the outburst, thus ruling out the influence of secondary stars. Moreover, the primary in the orbital phase of 0.5-0.9 is moving away from us, which cannot be caused by (2).

The accretion flow from the secondary star collides with one side of the accretion disk to form the hot spots, which are brightest when the hot spots face us at the orbital phase of about 0.7-0.9 \citep{warner1995cat}.
\citet{robinson1979quasi} observed that the QPOs was still found in the DN U Gem during the eclipse of the hot spots, thus ruling out the origin of QPOs from the hot spots. In addition, the orbital hump caused by the hot spots is not significant with the brightening of the accretion disc.
Although \citet{robinson1979quasi} ruled out the origin of QPOs from hot spots, they did not analyze the influence of the hot spots on the oscillation intensity of QPOs. In addition, the amplitude of QPOs of U Gem can be found in \citet{robinson1979quasi}'s work to be significantly larger around about orbital phase 0.9 than elsewhere (see \citet{robinson1979quasi} for details). 
Combined with our analysis, we believe that although hot spots are excluded from the origin of QPOs and are not significant at the time of outburst, it is still uncertain whether they contribute to the intensity of QPOs. We suggest that the relationship between the intensity of QPOs and the orbital phase has a high research value and can be used as a window to study the origin of QPOs.


\section{CONCLUSIONS} \label{sec:CONCLUSIONS}

In this paper, based on the data observed by TESS for HS 2325+8205, we have conducted in-depth research on the peculiar light curves and the QPOs. 
Firstly, we corrected the orbital period of HS 2325+8205 as 0.19433475(6)d using LSP. We obtained 491 new mid-eclipse times using Gaussian fit.
Secondly, we found that HS 2325+8205 has four different types of outbursts, but there are two that may not be actual, and the outburst period is about 13.83 days. In addition, there are double humps in quiescence, which may be caused by ellipsoidal modulation.
Finally, based on the LSP, CWT and sine fitting methods, we found for the first time that there are QPOs with a period of $\sim 2160$ s in the long outburst of HS 2325+8205. In addition, we also found that the intensity of QPOs is related to the orbital phase. 
The intensity of QPOs in orbital phases 0.5-0.9 are stronger than in orbital phases 0.1-0.5.
Through the analysis of orbital signals, 
we excluded the effects of the irradiation of the secondary and the ellipsoidal modulation, but the effect of hot spots on the intensity of QPOs needs to be further investigated.

\section*{Acknowledgements}

This work was supported by the National Natural Science Foundation of China (Nos. 11933008, U1831120, U1731238).
This work has made use of the data collected by the TESS mission, which is publicly available at the Mikulski Archive for Space Telescopes (MAST) and the Transiting Exoplanet catalogue (TEPcat) data.

\section*{Data Availability}

The TESS data used in this work can be obtained from the Mikulski Archive for Space Telescopes (MAST)\footnote{https://mast.stsci.edu/} and
the ExoFOP-TESS webpage\footnote{https://exofop.ipac.caltech.edu/tess/target.php?id=172518755}, respectively. 
We provide all mid-eclipse times of HS 2325+8205 from TESS online in the supplementary materials.
The corresponding author will share all other data underlying this article on reasonable request.



\bibliographystyle{mnras}
\bibliography{example} 

\bsp	
\label{lastpage}
\end{document}